\documentclass[fleqn,twoside,twocolumn,nofootinbib,showkeys]{revtex4} 
\usepackage[nocpr,nopacs]{ujp_UTF8} 

\usepackage{amsthm}

\begin{document}

\title[The $\mu$-Deformed Einstein Field Equations]%
{THE $\boldsymbol\mu$-DEFORMED EINSTEIN\\ FIELD EQUATIONS WITH
$\boldsymbol\mu$-DEPENDENT\\ EFFECTIVE COSMOLOGICAL CONSTANT}
\author{O.P.~Mykhailiv, Yu.A.~Mishchenko, A.M.~Gavrilik}
\affiliation{Bogolyubov Institute for Theoretical Physics, Nat.
Acad. of Sci. of Ukraine}
\address{14-b, Metrolohichna Str., Kyiv 03143,
Ukraine}%
\email{mykhailiv@bitp.kyiv.ua, mishchenko@bitp.kyiv.ua,
omgavr@bitp.kyiv.ua}

\autorcol{O.P.\hspace*{0.7mm}Mykhailiv, Yu.A.\hspace*{0.7mm}Mishchenko,
    A.M.\hspace*{0.7mm}Gavrilik}%

\setcounter{page}{831}

\begin{abstract}
In this paper, we  derive the $\mu$-deformed Einstein field
equations from the generalized thermodynamic functions of the
$\mu$-deformed analog of Bose gas model, applying the (adapted)
Verlinde's approach.\,\,The basic role of deformation parameter is
shown: it provides the possibility to vary the value of the
cosmological constant.\,\,Due to this, we suggest an interesting
treatment of the cosmological constant (CC) problem within the
framework of $\mu$-deformation.\,\,Na\-mely, viewing the derived
$\mu$-deformed CC as an effective one and varying the parameter
$\mu$ appropriately, we gain the possibility to drastically reduce
the CC, so as to get, for it, the realistic value.\,\,The relation
to dark matter is of importance.
\end{abstract}

 \maketitle

\section{Introduction}

Various deformed algebras have provided valuable insights into a
wide range of problems across diverse branches of physics.\,\,Among
different types of deformation, which are most conveniently and
clearly characterized by the deformation structure function (DSF),
see, e.g., \cite{1}, there are very popular and well-studied
deformed algebras of the exponential type like the $q$-oscillators
\cite{2,3,3a} or $p,q$-oscillators \cite{4}.\,\,Un\-like, the
$\mu$-deformation first introduced by Jannussis \cite{5} belongs to
a very different class of noncanonical Heisenberg algebras, namely,
the class of rational deformations.\,\,That implies very different
properties, and non-Fibonacci nature \cite{6} is one of
them.\,\,This extension has opened new perspectives for exploring
quantum systems with modified algebraic structures, offering
potential applications in such fields as high-energy physics,
quantum gravity, and physics of dwarf galaxies.

The $\mu$-deformation was developed in several aspects and applied
to various physical problems: the quasi-Fibonacci nature of the
nonlinear $\mu$-deformed oscillator has been established, as well as
for the proposed extensions or hybrid cases $(\mu; p, q)$ \cite{6};
the application to deformed versions of the non-relativistic Bose
gas model has been constructed \cite{7,9,10}; the intercepts of
$r$-particle momentum correlation functions in the $\mu$-Bose gas
model have been derived \cite{7,8}.\,\,Clo\-sely related
deformations have been introduced and shown as able to account for
compositeness and interaction of particles \cite{11}.\,\,Relevant
version of deformation was successfully applied to describe the
entanglement entropy of composite (quasi)boson systems \cite{12,13};
it is worth noting the temperature dependence of virial coefficients
and correlation function intercepts in the $(\mu, q)$-Bose gas model
\cite{14,15}, the condensate of the $\mu$-Bose gas as an effective
model of galactic-halos dark matter \cite{16,17}, the halo density
profile of dwarf or low surface brightness galaxies and their
rotation curves~\cite{18}.
Also, one should mention the related research on deformed versions
of Heisenberg algebra for the position and momentum operators.\,\,In
particular, the three-parametric $(p, q, \mu)$-deformed Heisenberg
algebra \cite{19} was shown to possess unusual properties, including
pseu\-do-her\-miticity of the involved operators.\,\,The special new
$\mu$-deformed Heisenberg algebra has been derived \cite{20} in the
context of the approach to dark matter based on the
$\mu$-deformation~\cite{18}.

Now, let us go over to one of well-known approaches to quantum
gravity, namely, the induced gravity theory.\,\,As known, it
branches into the Sakharov's quantum-field approach \cite{21,22}, on
one hand, and the emergent (or thermostatistical, or entropic)
theory on the other.\,\,In general context of quantum gravity,
Verlinde formulated a theory of induced gravity, wherein gravity is
understood as an emergent phenomenon arising from entropic forces
\cite{23}.\,\,His approach is rooted in statistical mechanics and
involves the holographic principle, as originally proposed by 't
Hooft \cite{24}; it is distinguished from Padmanabhan's
thermostatistical perspective on gravity~\cite{25}.

Verlinde's framework postulates that gravitational dynamics, as
described by either Newtonian theory or General Relativity, are not
fundamental interactions but emerge as effective macroscopic
description from an underlying statistical system.\,\,This system is
governed by classical statistics and holographic principle, where
the thermodynamic quantities, such as entropy, play a pivotal role
in the emergence of spacetime geometry and gravitational
forces.\,\,The entropic force, in this view, can be interpreted as a
manifestation of the underlying microscopic degrees of freedom
encoded in the statistical properties of the system~\cite{23}.

Furthermore, Verlinde's approach suggests that spacetime itself may
be an emergent phenomenon, derived from the collective behavior of
microscopic degrees of freedom, much like thermodynamic quantities
emerge from the microscopic states of matter.\,\,This perspective
has profound implications for the understanding of quantum gravity,
providing a potential bridge between the thermodynamic description
of space-time and the quantum mechanical description of fundamental
interactions.\,\,The holographic principle further strengthens this
connection, implying that the degrees of freedom (bits of
information) describing the gravitational system are encoded on a
lower-dimensional boundary, akin to the AdS/CFT correspondence in
string theory \cite{53,54}.\,\,Thus, Verlinde's theory opens new
avenues for exploring the relationship between gravity,
thermodynamics, and quantum mechanics.

In addition to classical statistics, deformed analogs of quantum
statistics can be applied in the context of induced gravity
\cite{31,32}.\,\,The corresponding physical systems (e.g., quantum
black holes) will be described by deformed thermostatistical
functions and, by applying the Verlinde's approach to them, one
obtains deformed analogs of the Einstein field
equations.\,\,Adop\-ting Ubriaco's one-parameter quantum group
$SU_{q}(2)$ interacting boson gas model results in the respective
$q$-deformed Einstein field equations \cite{26}, the use of $(p,
q)$-deformed Fermi gas model induces  $(p,q)$-deformed Einstein
field equations \cite{27,28}, and for
Viswanathan--Parthasarathy--Jagannathan--Chaichian (VPJC)
$q$-deformed fermion gas model there arises  $q$-deformed Einstein
field equations \cite{29}; likewise, with $q$-deformed fermion gas
model in two-dimensional space yet another type of $q$-deformed
Einstein field equations does emerge~\cite{30}.

Deformed Einstein equations for the $q$-deformed boson and
$q$-fermion gas models at the high-tem\-pe\-ra\-ture limit are
studied in Ref.~\cite{33}.\,\,At last, in Ref.~\cite{36}, two
versions of modified Einstein equations were obtained basing on the
GUP corrected Unruh temperature and the Verlinde's
approach.\,\,Verlinde's approach has encountered an ambiguous
reflections and some criticism, see, e.g.,
Ref.~\cite{37,38,39,40,41,42,43,44}.\,\,On the other hand, there is
a wide range of modified en\-tro\-pic gravities, that includes
noncommutativity, un\-gra\-vi\-ty as conformal invariant fields,
asymptotically safe gravity, Debye energy correction, surface
entropic gra\-vi\-ty, \textit{etc.}~\cite{44,45,46,47,48,52}.

The paper is organized as follows.\,\,First, a brief review of the
$\mu$-Bose thermodynamics is given.\,\,Se\-cond, we will derive a
$\mu$-deformed analog of the Einstein equation following the logic
of Verlinde's approach and basing on the $\mu$-deformed Bose gas
model.\,\,Note that, within the developed version of entropic
gravity, the $\mu$-deformed Friedmann equation can be inferred and
its application to cosmology can be explored.\,\,Third, the
implications for the obtained effective (depending on $\mu$)
extension of cosmological constant are analyzed in detail. The work
is ended with concluding remarks.\,\,Throu\-ghout the paper, we set
the units so that $c = \hbar = 1$.

\section{The $\boldsymbol\mu$-Deformed Einstein\\ Equations Using Verlinde Approach}

\subsection{The thermostatics of $\boldsymbol\mu$-Bose gas model}

We start with recalling some facts, which are necessary   for what
follows \cite{9,10}.\,\,In the standard Bose gas model, the total
number of particles is expressed as the Euler derivative of the
logarithm of the grand partition function $\mathcal{Z}$, i.e\vspace*{-1mm}
\begin{equation}
\mathcal{N} = z\frac{d}{dz} \ln \mathcal{Z},
\end{equation}
where the fugacity $z$ as function of the chemical potential
$\tilde{\mu}$ is given by $z = e^{\beta\tilde{\mu}}$, $\beta =
\frac{1}{k_{\rm B}T}$ is the inverse temperature involving the
Boltzmann constant $k_{\rm B}$.\,\,The grand canonical partition
function $\mathcal{Z}$ is treated through its logarithm by the
following expression:\vspace*{-2mm}
\begin{equation}
\ln\mathcal{Z} = - \displaystyle \sum_{i} \ln(1 - ze^{- \beta
\epsilon_{i}}),
        \label{lnZ}
\end{equation}
with $\epsilon_{i}$ being\vspace*{-1mm}
\begin{equation}
\epsilon_i = \frac{{|\bf p|}^2}{2m} = \frac{p_i^2}{2m}.
\end{equation}
From now on, we restrict ourselves to three-di\-men\-sional
space.\,\,To formulate the thermodynamic framework for the
$\mu$-analog of the Bose gas model, the standard expression for the
total number of particles $\mathcal{N}$ is supposed to incorporate
the $\mu$-dependence.\,\,Na\-mely, the modified definition for the
total number of particles is given as\vspace*{-2mm}
\begin{equation}
\mathcal{N}^{(\mu)} = z \, \mathcal{D}^{(\mu)}_z\ln \mathcal{Z} = -z
\, \mathcal{D}^{(\mu)}_z \sum_i \ln \left(\! 1 - z e^{-\beta
\epsilon_i} \!\right)\!,
\end{equation}
where we use the $\mu$-derivative\vspace*{-1mm}
\begin{equation}
\mathcal{D}^{(\mu)}_{x}f(x) = \int\limits_0^1 f'_{x} (t^{\mu}x) \,
dt, \quad f'_{x}(t^{\mu}x) = \frac{d f(t^{\mu}x)}{dx},
\end{equation}
which acts on the monomials as\vspace*{-1mm}
\begin{equation}
\mathcal{D}^{(\mu)}_{x} x^{n} = [n]_\mu x^{n-1},
\end{equation}\vspace*{-5mm}

\noindent
where\vspace*{-2mm}
\begin{equation}
[n]_\mu \equiv \frac{n}{1 + \mu n}, \quad 0 \leq\mu \leq1.
\end{equation}
Then the  $\mu$-deformed operator $\mathcal{D}^{(\mu)}$ is applied
to the logarithm of the grand canonical partition function  in
Eq.~(\ref{lnZ}), and the modified total number of particles
reads\vspace*{-3mm}
\[\mathcal{N}^{(\mu)} = z \sum_i \sum_{n=1}^{\infty}
        e^{-\beta \epsilon_i n} \frac{[n]_\mu}{n} z^{n-1}
= \]\vspace*{-6mm}
\begin{equation}
=\sum_i \sum_{n=1}^{\infty} \frac{[n]_\mu}{n}
                    \left( e^{-\beta \epsilon_i} \right)^n z^n.
                            \label{N(mu)}
\end{equation}
For the series to converge, the following condition for the product
$e^{-\beta \epsilon_i }z$ must hold:
\begin{equation}
\lim_{n \rightarrow \infty}\sum_i
\sum_{n=1}^{\infty}\frac{[n]_\mu}{n} \left(\! e^{-\beta \epsilon_i}
\!\right)^{\!n} z^n {<\infty} \Rightarrow
    0 \leq |z e^{-\beta \epsilon_i}| < 1.
\end{equation}
Separate the contribution of the $p_{i}=0, i = 0$ term from the
remaining sum.\,\,This yields:
\begin{equation}
\mathcal{N}^{(\mu)} = {\sum_i}' \sum_{n=1}^{\infty}
\frac{[n]_\mu}{n} \left(\!e^{-\beta \epsilon_i}\!\right)^{\!n} z^n +
    \sum_{n=1}^{\infty} \frac{[n]_\mu}{n} z^n.
            \label{N(mu)-2}
\end{equation}
As is well known, for a large volume $V$ and a large number of
particles $\mathcal{N}$, the spectrum of single-particle states
becomes nearly continuous.\,\,Due to this, we replace the summation
in Eq.~(\ref{N(mu)}) by a 3-integral over the 3-momentum space
\begin{equation}
\sum_i \rightarrow \frac{V}{(2\pi \hbar)^3} \int d^3k.
\end{equation}
Then we obtain
\begin{equation}
\mathcal{N}^{(\mu)} = \frac{V}{\lambda_{T}^3} \sum_{n=1}^{\infty}
\frac{[n]_\mu }{n^{5/2}}z^n + \mathcal{N}^{(\mu)}_0, \quad
\mathcal{N}^{(\mu)}_0 \equiv \sum_{n=1}^{\infty} \frac{[n]_\mu}{n} z^n.
\end{equation}
Here the thermal wavelength $\lambda_{T}$ is defined in terms of the
particle mass $m$ and the temperature $T$ (viewed as the
thermodynamic temperature):
\begin{equation}
\lambda_{T} = \sqrt{\frac{2 \pi \hbar^2}{m k_{\rm B} T}}.
\end{equation}
The standard Bose--Einstein function is given by the
\mbox{series}\vspace*{-3mm}
\begin{equation}
g_\ell(z) = \sum_{n=1}^{\infty} \frac{z^n}{n^\ell}.
\end{equation}
As a generalization of the latter, the $\mu$-polylogarithm is introduced, namely,
\begin{equation}
g^{(\mu)}_\ell(z) = \sum_{n=1}^{\infty} \frac{[n]_{\mu}}{n^{\ell+1}} z^{n}.
\end{equation}
Then the Eq.~(\ref{N(mu)-2}) becomes
\begin{equation}
\mathcal{N}^{(\mu)} = \frac{V}{\lambda_{T}^3} g^{(\mu)}_{3/2}(z) +
g^{(\mu)}_0(z),
\end{equation}
where\vspace*{-3mm}
\begin{equation}
g^{(\mu)}_{3/2}(z) = \sum_{n=1}^{\infty} \frac{[n]_\mu
}{n^{5/2}}z^n, \quad g^{(\mu)}_0(z) = \sum_{n=1}^{\infty}
\frac{[n]_\mu}{n} z^n.
\end{equation}
The $\mu$-deformed $\ln \mathcal{Z}^{(\mu)}$ is obtained by applying
the inverse of the Euler derivative, namely,
\begin{equation}
\ln \mathcal{Z}^{(\mu)} = \left(\! z \frac{d}{dz} \!\right)^{\!\!-1}
\mathcal{N}^{(\mu)}.
\end{equation}
After inserting $\mathcal{N}^{(\mu)}$
\[ \left(\! z \frac{d}{dz} \!\right)^{\!\!-1} \left(\! \frac{V}{\lambda_{T}^3}
    \sum_{n=1}^{\infty} \frac{[n]_{\mu}}{n^{5/2}}  z^n +
    \sum_{n=1}^{\infty} \frac{[n]_{\mu}}{n} z^n \!\right)= \]\vspace*{-6mm}
\begin{equation}
 =\frac{V}{\lambda_{T}^3} \sum_{n=1}^{\infty} \frac{[n]_{\mu}}{n^{5/2}}
 \left(\! z \frac{d}{dz}\! \right)^{\!\!-1}\!\! z^n +
 \sum_{n=1}^{\infty} \frac{[n]_{\mu}}{n}\left(\! z \frac{d}{dz}\!
 \right)^{\!\!-1}\!\!
 z^n,\!\!\!\!\!
\end{equation}
we get $\ln \mathcal{Z}^{(\mu)}$ in the form
\begin{equation}
\ln \mathcal{Z}^{(\mu)} = \frac{V}{\lambda_{T}^3} g^{(\mu)}_{5/2} + g^{(\mu)}_{1}.
\end{equation}
The internal energy $\mathcal{U}^{(\mu)}$ is determined through the
relation\vspace*{-3mm}
\begin{equation}
\mathcal{U}^{(\mu)} = -\left(\! \frac{\partial}{\partial \beta}
    \ln \mathcal{Z}^{(\mu)}\!\right)_{\!\!z, V}\!\!.
\end{equation}
With the usual $\beta$-derivative, we obtain the expression for
the $\mu$-deformed internal energy
\begin{equation}
-\frac{\partial}{\partial \beta}\left(\!\frac{V}{\lambda_{T}^3}
g^{(\mu)}_{5/2}
    + g^{(\mu)}_{1}\!\right)_{\!\!z, V}
\Rightarrow\ \mathcal{U}^{(\mu)} = - \tilde{\mu} \mathcal{N}^{(\mu)}.
\end{equation}
The equation of state then reads
\begin{equation}
\frac{\left(PV\right)_\mu}{k_{\rm B} T} = \ln \mathcal{Z}^{(\mu)} =
    \frac{V}{\lambda_{T}^3} g_{5/2}^{(\mu)}(z) + g_{1}^{(\mu)}(z).
            \label{EOS[PV]}
\end{equation}
The Helmholtz free energy of the model results as
\begin{equation}
\mathcal{H}^{(\mu)} = \tilde{\mu} \mathcal{N}^{(\mu)} -
\left(PV\right)_\mu\!.
            \label{H[PV]}
\end{equation}
The $\mu$-deformed entropy  can be obtained from the relation
$\mathcal{S}^{(\mu)} =
\frac{1}{T_\mathrm{U}}\left(\mathcal{U}^{(\mu)} -
\mathcal{H}^{(\mu)}\!\right)$
 (with $T_\mathrm{U}$ viewed as the \textit{Unruh temperature} \cite{54}) to yield
\[ \mathcal{S}^{(\mu)} = \frac{1}{T_\mathrm{U}} \biggl(\!
-2\tilde{\mu}\frac{V}{\lambda_{T}^{3}}g^{(\mu)}_{3/2}(z) -
2\tilde{\mu}g^{(\mu)}_{0}(z) + E\frac{V}{\lambda_{T}^{3}}\,
\times\]\vspace*{-7mm}
\begin{equation}
\times \, g^{(\mu)}_{5/2}(z)+ Eg^{(\mu)}_{1}(z)\!\biggr)\!.
                    \label{S(mu)}
\end{equation}
In the $\mu$-deformed extension of Verlinde's emergent gravity, we
use the identification $E = k_{\rm B}T$  rather than
$\frac{1}{2}k_{\rm B}T$, assuming each $\mu$-boson encodes two
fundamental bits of information.\,\,Phy\-sically, $\mu$-bosons may
be viewed as composite Bose-like particles built of correlated or
entangled pairs of elementary constituents (of either Fermi or Bose
type, see~\cite{11,12,13,14,15} for the former
one).\,\,Con\-sequently, each $\mu$-boson carries twice the energy
of a single bit in standard equipartition, leading to the total
energy $E=$ $=N k_{\rm B} T$ and preserving thermodynamic
consistency when $\mu$-deformation modifies the entropy-energy
relation along with the emergent gravitational
coupling.\,\,Exp\-licitly rewriting the expression for the thermal
wavelength and chemical potential in terms of $E = k_{\rm B}T$,
\begin{equation}
\lambda_{T}^{3} =
\left(\!\frac{2\pi\hbar^{2}}{mE}\!\right)^{\!\!3/2}\!\!, \quad
\frac{\tilde{\mu}}{E} = \ln z,
\end{equation}
we get the general expression for $\mu$-deformed entropy as function of $E$:
\[\mathcal{S}^{(\mu)} =
\frac{1}{T_U} \biggl(\!-2V
\left(\!\frac{m}{2\pi\hbar^{2}}\!\right)^{\!\!3/2}
    g^{(\mu)}_{3/2}(z)E^{5/2} \ln z- 2E \, \times\]\vspace*{-7mm}
\begin{equation}
\times \, g^{(\mu)}_{0}(z)\ln z +
V\left(\!\frac{m}{2\pi\hbar^{2}}\!\right)^{\!\!3/2}
    g^{(\mu)}_{5/2}(z)E^{5/2} + Eg^{(\mu)}_{1}(z)\!\biggr)\!.
\end{equation}
This is of principal importance for what follows.

\subsection{Derivation of the \boldmath$\mu$-deformed\\ Einstein equations
            using Verlinde approach}

In Verlinde's approach to entropic gravity, the system approaches to
statistical equilibrium, when the entropic force stemming from the
changes in entropy, becomes balanced with  the forces that
contribute to an increase in the entropy.\,\,Ac\-cor\-dingly, the
formulation suggests that gravity itself can be understood as an
emergent force arising from the entropic dynamics of underlying
microscopic degrees of freedom, rather than a fundamental
interaction.\,\,At equilibrium, the total entropy $\mathcal{S}$ of
the system remains constant, and it reaches an extreme
value.\,\,This is a direct consequence of the second law of
thermodynamics, which states that entropy tends to increase until a
maximum is reached under the given constraints.\,\,In what follows,
the entropy $\mathcal{S}^{(\mu)}(E, x^\nu)$ satisfies:
\begin{equation}
\frac{d}{d x^\nu}
\mathcal{S}^{(\mu)}(E, x^\nu) = 0.
\end{equation}
From Verlinde's perspective, the system's equilibrium is not only a condition
of the force balance, but also a reflection of the deeper thermodynamic
nature of gravity
\begin{equation}
\frac{\partial \mathcal{S}^{(\mu)}}{\partial E} \frac{\partial
E}{\partial x^\nu} + \frac{\partial \mathcal{S}^{(\mu)}}{\partial
x^\nu} = 0,
            \label{VerlP}
\end{equation}
where the components represent the energy gradient and an entropic force
\begin{equation}
\frac{\partial E}{\partial x^\nu} = -\mathcal{F}_\nu, \quad
\frac{\partial \mathcal{S}}{\partial x^\nu} = \nabla_\nu \mathcal{S}.
            \label{dE/dx}
\end{equation}
From the derivative of $\mathcal{S}^{(\mu)}$, we have
\[ \frac{\partial \mathcal{S}^{(\mu)}} {\partial E} =
\frac{1}{T_U} \biggl(\!\frac{5V}{2}g^{(\mu)}_{5/2}(z)
\left(\!\frac{mE}{2\pi\hbar^{2}}\!\right)^{\!\!3/2}-
5Vg^{(\mu)}_{3/2}(z) \, \times\]\vspace*{-6mm}
\begin{equation}
\times \left(\!\frac{mE}{2\pi\hbar^{2}}\!\right)^{\!\!3/2}\ln z
    + g^{(\mu)}_{1}(z) - 2g^{(\mu)}_{0}(z)\ln z \!\biggl)\!.
\end{equation}
For simplicity, let us use the notation
\begin{align}
&G_1(\mu;z) =
\frac52 \frac{V}{\lambda_T^3}
        \left(\! g^{(\mu)}_{5/2}(z)  -  2\ln z g^{(\mu)}_{3/2}(z)\! \right)\!,
                                                            \label{Def_G1}\\
&G_2(\mu;z) = g^{(\mu)}_1(z) - 2g^{(\mu)}_0(z) \ln z.
                                                            \label{Def_G2}
\end{align}
The formula for the derivative then takes the form
\begin{equation}
\frac{\partial \mathcal{S}^{(\mu)}} {\partial E} =
    \frac{1}{T_\mathrm{U}}\Bigl(\! G_{1}(\mu;z) + G_{2}(\mu;z)\!\Bigr)\!.
                \label{dS/dE}
\end{equation}
By inserting formulae (\ref{dS/dE}) and (\ref{dE/dx}) in
Eq.~(\ref{VerlP}), we obtain the relation
\begin{equation}
\left(G_{1}(\mu;z) + G_{2}(\mu;z)\right)\mathcal{F}_\nu
        = T_\mathrm{U}\nabla_\nu \mathcal{S}.
\end{equation}
Here the entropic force $\mathcal{F}_\nu = - m e^{\varphi}
\nabla_\nu \varphi$ and $\varphi$ is a GR generalization of Newton's
potential:
\begin{equation}
\varphi = \frac{1}{2} \log(-\xi^{\nu}\xi_{\nu}).
\end{equation}
The change of entropy through the holographic screen resulting from
a displacement of particle (bit) by one thermal wavelength
$\lambda_{T}$ along some normal direction ${\bf n}$ to the screen
determined by $n_\nu (\nu=0,1,2,3)$ reads
\begin{equation}
\nabla_\nu \mathcal{S} = - 2\pi\frac{m}{\hbar} n_\nu.
\end{equation}
In Verlinde's emergent gravity framework, the derivation of
Einstein's field equations is performed in thermodynamic terms by
employing the holographic principle, the Unruh effect, and the
Bekenstein entropy bound, all of which can be naturally interpreted
within the AdS/CFT correspondence.\,\,Hence, the accelerating
observer may be regarded as intrinsically coupled to the framework
of induced gravity, since the Unruh temperature  perceived by such
an observer constitutes a macroscopic manifestation of the same
underlying microscopic degrees of freedom whose collective dynamics
gives rise to the emergent gravitational field.\,\,Then (on equating
the thermodynamic temperature to the Unruh temperature), we obtain
the relation that involves the $\mu$-deformed temperature of the
form:\vspace*{-2mm}
\begin{equation}
T_\mathrm{U} = (G_{1}(\mu;z) + G_{2}(\mu;z))e^{\varphi}
    \nabla_\nu \varphi \frac{\hbar}{2\pi}n_\nu.
                \label{T_U}
\end{equation}
With a screen located on a closed surface $\mathcal{\tilde{S}}$
at a ``constant redshift'' the Komar mass can be written as
\begin{equation}
\mathcal{M} = \frac{1}{2} \int\limits_{\mathcal{\tilde{S}}}
T_\mathrm{U} d\mathcal{N}.
            \label{KomM-0}
\end{equation}
The infinitesimal change  of the number of bits can be written,
according to Bekenstein idea~\cite{31}, as
\begin{equation}
d\mathcal{N} = \frac{d\mathcal{A}}{G\hbar}.
                \label{dN}
\end{equation}
By using Eq.~(\ref{dN}) and Eq.~(\ref{T_U}) in Eq.~(\ref{KomM-0}),
we obtain the total mass
\begin{equation}
\mathcal{M} = \frac{1}{4\pi G} \int\limits_{\mathcal{\tilde{S}}}
\mathcal{G}_V(\mu;z)e^{\varphi}  \nabla_\nu \varphi
n_\nu\,d\mathcal{A},
\end{equation}
where
\begin{equation}
\mathcal{G}_V(\mu;z) \equiv G_{1}(\mu;z) + G_{2}(\mu;z).
\end{equation}
Next, from the Stokes' theorem, the Killing equation, and the relation
$\nabla_\nu \nabla^\nu \xi^{\mu} = -\mathcal{R}_{\nu}^{\mu} \xi^{\nu}$,
we obtain the desired $\mu$-dependent total mass in the form
\begin{equation}
\mathcal{M} =  \frac{1}{4\pi G} \int\limits_V \mathcal{G}_V(\mu;z)
    n^{\nu} \xi^{\mu} \, \mathcal{R}_{\mu\nu} \, dV.
            \label{KomM-1}
\end{equation}
On the other hand, there is an alternative expression for the Komar
mass in terms of energy-momentum tensor:\vspace*{-2mm}
\begin{equation}
\mathcal{M} = 2 \int\limits_V \left(\! \mathcal{T}_{\mu\nu} -
      \frac{1}{2} g_{\mu\nu} \mathcal{T} + \frac{\Lambda}{8\pi G}g_{\mu\nu}
    \!\right) n^{\nu} \xi^{\mu} \, dV.
                \label{KomM-2}
\end{equation}
In this framework, the $\mu$-deformed mass refers to the
thermodynamic energy of a $\mu$-Bose gas distributed across the
holographic volume.\,\,The Komar mass, by contrast, is an
$\mu$-deformed emergent gravitational quantity that inherits
deformation effects from the underlying thermodynamic system via the
AdS/CFT correspondence.\,\,Then from Eqs.~(\ref{KomM-2}),
(\ref{KomM-1}), we infer
\[2 \int\limits_V \left(\! \mathcal{T}_{\mu\nu} - \frac{1}{2} g_{\mu\nu} \mathcal{T}
+ \frac{\Lambda}{8\pi G}g_{\mu\nu}\! \right) n^{\nu} \xi^{\mu} \, dV
= \]\vspace*{-7mm}
\begin{equation}
=\frac{\mathcal{G}_V(\mu;z)}{4\pi G}
    \int\limits_V \mathcal{R}_{\mu\nu} n^{\nu} \xi^{\mu} \,   dV.
\end{equation}
From this we obtain the $\mu$-deformed Einstein equations describing
$\mu$-induced gravity acting on deformed matter fields
\begin{equation}
2\left(\!\mathcal{T}_{\mu\nu} - \frac{1}{2} g_{\mu\nu} \mathcal{T}
    + \frac{\Lambda}{8\pi G}g_{\mu\nu}\!\right)
    = \frac{\mathcal{G}_V(\mu;z)}{4\pi G}\mathcal{R}_{\mu\nu}.
            \label{Ruv}
\end{equation}
At last, taking the trace of Eq.~(\ref{Ruv}), yields the desired
$\mu$-deformed Einstein equation
\begin{equation}
\mathcal{R}_{\mu\nu} - \frac{1}{2}g_{\mu\nu}\mathcal{R} +
\frac{\Lambda}{\mathcal{G}_V(\mu;z)}g_{\mu\nu} =
    \frac{8\pi G}{\mathcal{G}_V(\mu;z)}\mathcal{T}_{\mu\nu}.
            \label{mu-Einst}
\end{equation}
As  seen in this equation, the modified (or\textit{ effective}
$\Lambda_\mathrm{eff} \equiv \frac{\Lambda}{\mathcal{G}_V(\mu;z)}$)
\textit{cosmological constant} (ECC) has appeared, along with
\textit{modified gravitational constant}.

In the both deformed ($\mu>0$) and undeformed ($\mu=0$) situations,
the ECC and also the gravitational constant (GC) in the resulting
$\mu$-deformed Einstein Eqs.\,\,depend on the \textit{variable}
fugacity $z$ of underlying inducing system.\,\,To get rid of $z$,
natural way is to fix $z$ as $z=1$ (or some value very close to
$1$).\,\,That implies dealing with Bose-like condensate.\,\,Now, the
crucial point is that the  $\mu =0$ (i.e.\,\,pure Bose gas) case,
for fixed $z = 1$, results in vanishing CC and GC, what is clearly
unphysical.\,\,Thus, pure Bose gas cannot serve as underlying (or
inducing) system while the ($\mu$-)deformed extension will be of
principal importance, as will be seen from the treatment below.

\section{Analysis of the Effective\\ Cosmological Constant}

We study the behavior of effective cosmological constant, with more
spectacular plotting of the factor $\mathcal{G}_V(\mu;z) \equiv
G_1(\mu;z)+G_2(\mu;z)$.\,\,{The functions $G_1(\mu;z)$ and
$G_2(\mu;z)$ defined by~(\ref{Def_G1})--(\ref{Def_G2}) involve slow
convergent $\mu$-polylogarithmic series $g^{(\mu)}_s(z) =$ $=
\sum_{n=1}^\infty (1 + \mu n)^{-1}\,\frac{z^n}{n^s}$, for $s= 0, 1,
\frac32, \frac52$,} see e.g. \cite{10} for a definition and some
uses.\,\,In view of calculation nontriviality of the
$g^{(\mu)}_s(z)$ series at $z$ close to $1$, and small~$\mu$, we
applied \textit{Riemann sum approximation} to appropriate terms of
$G_2(\mu;z)$ described below, as well as \textit{Euler--Maclaurin
formula} \cite{59}  when treating~$G_1(\mu;z)$.\,\,Be\-sides, we use
the expansion of polylogarithm $\operatorname{Li}_s(z)$  in $(\ln
z)^n$
\begin{equation}
\begin{array}{l}
\displaystyle\operatorname{Li}_s(z) =
\Gamma(1-s)\,\bigg(\!\ln{\frac{1}{z}} \!\bigg)^{\!\!s-1}\!\! +
       \!\! \sum\limits_{n=0}^\infty\! \zeta (s-n) \frac{(\ln{z})^n}{n!},
        \\[5mm]
s\neq 1,2,3,...,\ |\ln z| < 2\pi, \quad  n=1,2,...
\end{array}\!\!\!\!\!\!\!\!\!\!\!\!
\end{equation}
as well as its integer order $s$ counterpart, see e.g. \cite{57} or
\cite{59,60}.

\subsection{Treatment of the
function~\boldmath$G_1(\mu;z)$}

\begin{figure}
\includegraphics[width=8cm]{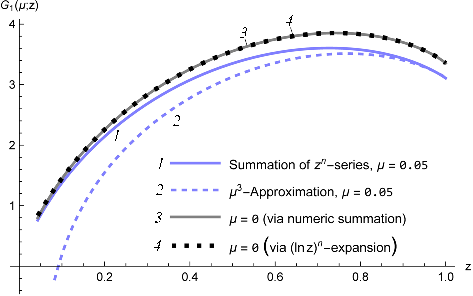}\\
{\it a}\\[2mm]
\includegraphics[width=8cm]{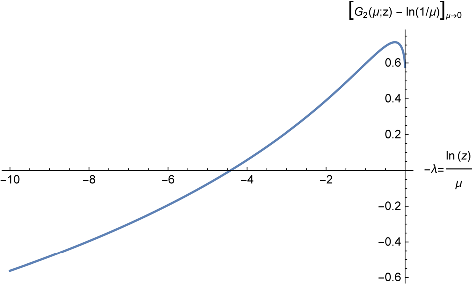}\\
{\it b}\\[2mm]
\includegraphics[width=8cm]{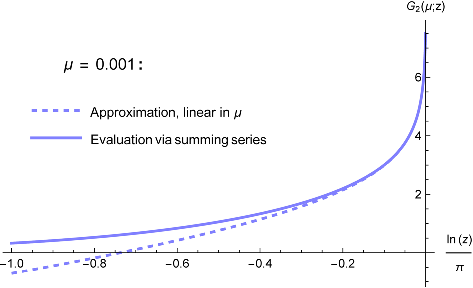}\\
{\it c} \vskip-3mm\caption{Dependence of~$G_1$ on~$z$ (\textit{a})
and the difference  $G_2 - \ln\frac1{\mu}$  on $-\lambda=\frac{\ln(z)}{\mu}$ (\textit{b}). The
bottom plots show a discrepancy between the approximation and
numerical evaluation for~$G_2$. Pane~(\textit{a}): solid
curves~\textit{1} and~\textit{3} show the behaviour of numerically
evaluated~$G_1$ basing on the summation of defining $z^n$-series,
resp. for $\mu=0.05$ and $\mu=0$; dashed~\textit{2} and
dotted~\textit{4} curves are obtained via $\mu^3$-approximation
(\ref{G1-Appr}) and $(\ln z)^n$ expansion (\ref{G1(0;z)}), resp. for
$\mu=0.05$ and $\mu=0$. Pane~(\textit{c}): dashed curve corresponds
to linear-in-$\mu$ approximation, see~(\ref{G2-Appr}) and
footnote~\ref{ft1}, whilst solid one is evaluated via summing series
(for both cases $\mu=0.001$)}   \label{fig1}
\end{figure}

We introduce the auxiliary notation $\Xi_T \equiv
\frac{V}{\lambda_T^3}$ here.\,\,Going over from the
fugacity~$z=e^{\beta\tilde{\mu}}$ to new reduced variable
\begin{equation}
\lambda =
\lambda(z,\mu)  = - \frac1{\mu} \ln z  \approx\frac{1-z}{\mu},
        \label{Def_la(z)}
\end{equation}
for $z$ close to~$1$, we arrive at the desired `small~$\mu$' approximation
\[
G_1(\mu;z) = \frac52\, \Xi_T \biggl\{\! \zeta \left(\!\frac52\!
\right) + \zeta\left(\!\frac32\!\right) (\lambda - 1)\,\mu \, -
            \]\vspace*{-6mm}
\[
- \, 2\sqrt{\pi} \biggl[ \! \left(\! \frac43\lambda - 1\!\right)
\lambda^\frac12 +
      (2\lambda - 1)\, \mathop{\textsf{M}_\mathrm{erfc}}( \sqrt{\lambda})
    \biggr]\, \mu^\frac32\, -
            \]\vspace*{-6mm}
            \[
-\, \frac12 \zeta \left(\!\frac12\!\right) (3\lambda-2)\lambda \mu^2
+ \frac{1}{3!} \zeta\left(\! -\frac12\!\right) (5\lambda-3)\lambda^2
\mu^3 \!\biggr\} + \]\vspace*{-6mm}
\begin{equation}
       + \, \lambda^{'} O\bigl(\mu^2 \lambda^{\prime2} \bigr),
                    \label{G1-Appr}
\end{equation}
where $\mu \to 0$, $\lambda \ll \frac{2\pi}{\mu}$, $\lambda^\prime
\equiv \lambda+1$ and \textit{Mills' ratio} for the complementary
error function
\begin{equation}
\operatorname{erfc}(x) \equiv 1 - \operatorname{erf}(x) \equiv
        { \frac2{\sqrt{\pi}}  \int\limits_x^\infty  e^{-t^2} dt },
\end{equation}
defined \cite{59,60} as
\begin{equation}
\mathop{\textsf{M}_\mathrm{erfc}}(x) =
    \frac{\int\nolimits_x^\infty e^{-t^2} dt}{e^{-x^2}}  \equiv
    \frac{\sqrt{\pi}}2\, e^{x^2} \operatorname{erfc}(x),
            \label{MillsRat}
\end{equation}
and $\zeta(x)$ is Riemann zeta function.\,\,The original fugacity
variable~$z$ is recovered as
\begin{equation}
 z = z(\lambda,\mu) = e^{-\mu \lambda}.
\end{equation}
To explore the behavior of~$G_1(\mu;z)$ and total
$\mathcal{G}_V(\mu;z)$ we set $\Xi_T = 1$  unless the contrary is
specified.\,\,For convenience and as evidence for continuous or
divergent behavior at~$z \to 1$, we also give the explicit
expressions for~$G_{1,2}(\mu;z)$ when $\mu = 0$, in terms of
$\tilde{\lambda} \equiv  -\ln (z)$,  $|\ln z| < 2\pi$:
\[
G_1(0;z) = \frac52\, \biggl(\! \zeta\bigg(\!\frac52 \!\bigg)  +
\zeta\bigg(\!\frac32\!\bigg) \tilde{\lambda} - \frac83\sqrt{\pi}\,
\tilde{\lambda}^{3/2}   - \frac12 \tilde{\lambda}^2\,
\times\]\vspace*{-6mm}
\begin{equation}
\times \sum\limits_{r=0}^\infty \,(-1)^{\big[\frac{r+1}2\big]}\,
\frac{ \zeta\bigl(r+1/2\bigr)}{r+1/2}\, \frac{ (2r +3)!!}{(r+2)!}
\bigg(\!\! -\frac{\tilde{\lambda}}{4\pi}
\!\bigg)^{\!\!r}\biggr)\!,\!\!\!\!\!\!
                        \label{G1(0;z)}
\end{equation}
where $[\frac{r+1}{2}]$ denotes the largest integer not exceeding
$\frac{r+1}{2}$, and $n!! = n(n-2)(n-4)\cdot...$ is double
factorial.\,\,Re\-mark that, as seen from {(\ref{G1-Appr}),
(\ref{G1(0;z)})}, in the close vicinity to $z=1$, where function
$G_{2}(0;z)$ is logarithmically divergent, the impact from
$G_{1}(\mu;z) \approx$ $\approx G_{1}(0;1) = \frac{5}{2}
\zeta(\frac{5}{2})$ for small $\mu$ is finite  (proportional to
$\Xi_T$).\,\,The precision of approximate formula~(\ref{G1-Appr})
and of expansion~(\ref{G1(0;z)}), compared to numerical evaluation
from the definitions, is demonstrated
in~Fig.~\ref{fig1},~\textit{a}.

\subsection{Treatment of the function~\boldmath$G_2(\mu;z)$}

Note that it is the function~{$G_2$} that makes much larger
contribution into the sum, if $1-\delta < z < 1$ and $\mu \to 0$.
For $G_2$, to derive analogous approximate formula, we first invoke
approximation of integer~$\mu^{-1}$ (which is reasonable due to
smallness of~$\mu$), that reads
\begin{align}
&g^{(\mu)}_0(z) = \mu^{-1} e^{\lambda(z,\mu)} \Biggl(\!
\operatorname{Li}_1(z)
 - \sum_{n=1}^M \frac{z^n}{n} \!\Biggr) - O(1),
\\
&g^{(\mu)}_1(z) = \operatorname{Li}_1(z) - \mu\, g^{(\mu)}_0(z),
\end{align}
for arbitrary continuous~$\mu$, $0<\mu<1$, with integer part~$M
\equiv [\mu^{-1}]$ of its inverse ratio.\,\,Then, to calculate the
sum~$\sum_{n=1}^M \frac{z^n}{n}$ explicitly (and approximately) we
apply the quadrature midpoint rule, namely
\[\sum_{n=1}^M \frac{1-z^n}{n} = \!\!\!
\int\limits_{\mu/2}^{(M+1/2)\mu}  \frac{1-e^{-\lambda x}\!}{x} dx\
-\ O_\lambda(\mu^2) = \]\vspace*{-6mm}
\begin{equation}
 =\operatorname{Ein}\bigg(\!\lambda + \frac12 \mu\lambda\!\bigg) -
  \operatorname{Ein}\bigg(\!\frac12 \mu\lambda\!\bigg) -
        2\, O\bigg(\!\mu + \frac{\mu^2\lambda^3}{72}\!\bigg)\!,
\end{equation}
with~{\cite{58,59}} modified exponential integral
\begin{equation}
\operatorname{Ein}(x) = \int\limits_0^x { \frac{1-e^{-t}}{t} } dt.
\label{Ein-Def}
\end{equation}
After some analysis we obtain\,\footnote{More detailed treatment
based on inverse-integer representation for~$\mu$
    yields formula with linear-in-$\mu$ correction:
    $
    G_2(\mu;z) = \ln\frac1{\mu} +
    \bigl[ (2\lambda - 1)\, e^\lambda\, E_1(\lambda) - \ln(\lambda) \bigr]
    - \frac12 (\lambda - 1)\, \mu\, +$ $+\, e^\lambda O(\mu^2 \lambda^{\prime4})
    $.\label{ft1}}
\[ G_2(\mu;z) = \ln\bigg( \!\frac1{\mu} \!\bigg) +  \big[
    (2\lambda - 1)\, e^\lambda\, E_1(\lambda) - \ln(\lambda)
\big]\, + \]\vspace*{-6mm}
\begin{equation}
+ \, e^{\lambda+2}\,  \bigg(\!
 O(\mu) +  \bigg[12 + \bigg(\!\frac{\delta z}{\mu}\!\bigg)^{\!\!2}  \bigg] O(\delta z)
\!\bigg)\!, \label{G2-Appr}
\end{equation}
with $\mu \to +0,\ \ \delta z \to +0$ and
\begin{equation}
\lambda \equiv  \lambda(z,\mu) =  \frac{ |\ln z|}{\mu}
 =\frac{\delta z}{\mu}  \bigg(\!1 + \frac{\delta z}2 + ...
 \!\bigg)\!,
\end{equation}
where $\delta z \equiv 1-z $ and ~$E_1(\lambda)$ denotes exponential
integral \cite{58,59}
\begin{equation}
E_{1}(x)=\int\limits_{x}^{\infty}\frac{e^{-t}}{t}dt, \quad
|\operatorname{arg}(x)|<\pi.
\end{equation}
Since, for the extremely small (but positive) values of $\mu$, the
function $G_1(\mu;z) \sim \mathrm{const}$, when $z$ is close to $1$,
the prevailing behavior of $G_1(\mu;z)+G_2(\mu;z)$ is determined by
the first meaningful terms of~$G_2$. The respective dependence on
$\lambda$ is shown in Fig.~\ref{fig1},~\textit{b}.\,\,The validity
of the approximation~(\ref{G2-Appr}) close to $z=1$ is demonstrated
in Fig.~\ref{fig1},~\textit{c}.

The {\textit{extremum condition for $\lambda$}}
\begin{equation}
(1 + 2\lambda) E_1(\lambda) = 2 e^{-\lambda}
\end{equation}
can be solved either analytically or numerically.\,\,The first
method involves known power series expansion for~$E_1(\lambda)$, see
e.g.~\cite{58},
\begin{equation}
E_1(\lambda) = -\gamma - \ln(\lambda) -
    \sum_{n=1}^\infty  \frac{(-1)^n \lambda^n}{n n!},
\end{equation}
where  $\gamma = 0.577215...$  is the Euler constant, implying
smallness of~$\lambda$, and leads for~$n\le 2$ to quadratic equation
\[
7.25 \lambda^2 -\bigl(1 + 4\ln 2 -2\gamma\bigr) \lambda +
\bigl(1 + \gamma -2\ln 2\bigr) = 0,
\]
with $\lambda \approx 0.2597...$\,.\,\,The second method provides
more precise solution
        $\lambda = \lambda^{(0)} = 0.2589...$,
found by means of \textit{Wolfram Mathematica software}.\,\,So
respective fugacity $z_0 \approx 1 - 0.26 \mu$. Corresponding
maximum value of $G_2(\mu;z)$ immediately follows
from~(\ref{G2-Appr}):\vspace*{-1mm}
\[
 \bar{G}_2 =  \ln\left(\! \frac1{\mu}\! \right)  +
  \lim\limits_{\mu \to 0}   \biggl[
    G_2\left(\!\mu, {z(\lambda^{(0)},\mu)}\!\right)
                        - \ln\left(\! \frac1{\mu} \!\right)
  \biggr]=\]\vspace*{-6mm}
\begin{equation}
=0.7159\,... + \ln\left(\! \frac1{\mu}\! \right)\!\! .
                \label{G2max}
\end{equation}
{\textit{\textbf{Remark.}}\,\,From the definition of parameter
$\lambda=$  \mbox{$=\lambda(z,\mu)$}, see~(\ref{Def_la(z)}), and the
known one for fugacity $z = e^{\beta\tilde{\mu}}$ we directly obtain
relation between chemical potential $\tilde{\mu}$ and deformation
parameter $\mu$:\vspace*{-1mm}
\begin{equation}
    \tilde{\mu} \equiv \beta^{-1}\ln z= - \mu\frac{\lambda(z,\mu)}{\beta}.
\end{equation}
So, for the case when $G_{2}(\mu;z)$ is prevailing over
$G_{1}(\mu;z)$ in the neighbourhood of $z = 1$, we  estimate
chemical potential for maximum point of $G_{1} + G_{2}$ (where
$\lambda(z_{0},\mu) \equiv \lambda^{(0)}$) as\vspace*{-1mm}
\begin{equation}
    \tilde{\mu}_{0}=-\lambda^{(0)}\mu k_{\rm B}T\approx  -0.26\mu \times k_{\rm B} T
\label{mu-Chem}
\end{equation}\vspace*{-5mm}

\noindent
with $\mu$ being small enough.

Explicit expression for~$G_2(\mu;z)$ at $\mu = 0$ and $|\ln z| <
2\pi$ is given as\vspace*{-3mm}
\[ G_2(0;z) = -\ln\left| \ln z \right| +  2 -
\frac12\,\tilde{\lambda}\! +\! \frac16\, \tilde{\lambda}^2
  \frac{ {}_1F_1(2;4;-\tilde{\lambda}) }{
  {}_1F_1(1;2;-\tilde{\lambda})}\, -\]\vspace*{-7mm}
\begin{equation}
-  \sum\limits_{m=2}^\infty\ B_m \frac{ \tilde{\lambda}^m }{m\,m!}
,\quad \tilde{\lambda} \equiv  -\ln (z). \label{G2(0;z)}
\end{equation}
Here  ${}_1F_1(...)$ is Kummer hypergeometric function and $B_m$ are
Bernoulli numbers, see e.g.\,\,\cite{58} for their
definitions.\vspace*{-1mm}

\subsection{The sum \boldmath$G_1+G_2$ at $z$ close to unity}

Let us also observe certain ``universality'' in dependence of
$G_1(\mu;z)+G_2(\mu;z)$ on~$z$: in  the  neighborhood of $z=1$ at
small $\mu$ it manifests itself through just combined variable
$\lambda(z,\mu)$ (as seen from (\ref{G1-Appr}) and
(\ref{G2-Appr})).\,\,Using this fact and the above approximations
for $G_1(\mu;z)$ and $G_2(\mu;z)$ we establish the following.

\textit{\textbf{Statement.}}\,\,The narrow, convex upwards,
maximum~$\bar{\mathcal{G}}_V(\mu)$ \mbox{\textit{{exists}}} for
arbitrarily small~$\mu$, $0 < \mu < \delta$, and is
\textit{{continuously}} tending to infinity ($+\infty$) when $\mu$
tends to zero, $\mu \to +0$.\,\,At the same time, the respective
fugacity, given by\vspace*{-2mm}
\begin{equation}
z_{0}\simeq 1 - \mu \lambda^{(0)} \approx 1 - \frac{\mu}{4},
\end{equation}
continuously tends from the left to unity ($\delta z_{0} \equiv 1\,
-$ $-\, z_{0} \simeq \frac{\mu}{4} \to +0$). The height of the
maximum, at small $\mu$, is given with `order of $\mu$' error
by\vspace*{-1mm}
\begin{equation}
\bar{\mathcal{G}}_V(\mu) \approx  0.7159... + \frac52\,
\zeta\bigg(\!\frac52\!\bigg)\Xi_T + \ln\bigg(\! \frac1{\mu} \!\bigg)
\label{G_Vmax}
\end{equation}\vspace*{-3mm}

\noindent and \textit{{is logarithmically increasing
with~$\frac{1}{\mu}$}}.

\textit{\textbf{Corollary.}}\,\,Taking the expression for
cosmological constant involved in $\mu$-deformed Einstein
equation~(\ref{mu-Einst}) as\vspace*{-3mm}
\begin{equation}
    \Lambda_\mu
    =  \Lambda_\mathrm{eff}(\mu;z)
    = \frac1{G_1(\mu;z) + G_2(\mu;z)} l_\mathrm{P}^{-2},
            \label{Def_La_mu}
\end{equation}\vspace*{-4mm}

\noindent where $l_\mathrm{P}$ is the Planck length, we find the
appropriate value~$\mu = \mu^{*}$ of deformation parameter to fit
the presently known value of cosmological constant\vspace*{-1mm}
\begin{align}
&\Lambda_{\rm ex}= 3 \biggl(\! \frac{H_0}{c} \!\biggl)^{\!\!2}
\Omega_{\Lambda}
    = 1.466 \times 10^{-52} m^{-2} =         \notag
\\[-1mm]
&=\underbrace{3.827 \times 10^{-123}}_{
                \Lambda_\mathrm{ex}^{(\mathrm{P})}}
                l_\mathrm{P}^{-2},
                    \label{Lambda_ex}
\end{align}\vspace*{-3mm}

\noindent where ``ex'' means observable.\,\,In view of extraordinary
smallness of $\Lambda_\mathrm{ex}^{(\mathrm{P})}$ we expect to
achieve the proper value (\ref{Lambda_ex}) of the effective
cosmological constant (\ref{Def_La_mu}) at the minimum point on the
curve~$\Lambda_\mu(z)$.\,\,So, to obtain this proper value for
$\Lambda_\mu$, we put\vspace*{-1mm}
\[
{\Lambda_\mu(z_0) = \frac1{\mathcal{G}_V(\mu;z_0)}\,
l_\mathrm{P}^{-2}
     =  \Lambda_\mathrm{ex}^{(\mathrm{P})}\,  l_\mathrm{P}^{-2} }
\]\vspace*{-4mm}

\noindent that is equivalent, see~(\ref{G_Vmax}), to\vspace*{-1mm}
\begin{equation}
\mathcal{G}_V(\mu;z_0) \approx
 0.72 + \frac52 \zeta\bigg(\!\frac52\!\bigg) +
            \ln\bigg(\! \frac1{\mu} \!\bigg)\! \bigg|_{\mu =
            \mu^\ast}\!\!\!
= \frac1{\Lambda_\mathrm{ex}^{(\mathrm{P})}}.\!\!\!\!\!
\label{G_V(mu*;z0)}
\end{equation}\vspace*{-3mm}

\noindent This yields the corresponding value of deformation
parameter~$\mu^\ast$, which provides \textit{{the above
value$~(\ref{Lambda_ex})$ in the extremum}},\vspace*{-1mm}
\begin{equation}
\mu^\ast \approx
e^{-1/\Lambda_\mathrm{ex}^{(\mathrm{P})} + 0.72 + 5/2\,\zeta(5/2)}
    \sim\; e^{-2.61 \times 10^{122}}\!.
\end{equation}\vspace*{-6mm}

\noindent In fact, the extremum at $z=z_0$ is so close to $z=1$
point on the $\mu=\mu^\ast$ curve that it is inessential what
fugacity from the $[z_0,\,1]$-interval is taken, resulting in almost
equal values~$\Lambda_\mathrm{eff}(\mu;z) \approx \Lambda_{\rm ex}$,
up to very high precision\,\footnote{
    Let us observe, using~(\ref{G1-Appr}) and (\ref{G2-Appr}),
    that the factor~$\mathcal{G}_V(\mu^\ast;z)$ taken at $z=1$ is lesser
    than  $\mathcal{G}_V(\mu^\ast;z_0) \sim 10^{122}$ in the extremum just
    by
    $\Delta\mathcal{G}_V(\mu^\ast) \approx 0.7159...\! - \gamma
        \approx 0.14 \ll 10^{122}$.
    So, due to~(\ref{Def_La_mu}), the effective cosmological constant
    remains almost unchanged.}.\,\,The fugacity~$z_0$ for this extremum stems from~(\ref{G2-Appr}):\vspace*{-2mm}
\begin{equation}
 z_0 = e^{-\mu \lambda^{(0)}} \approx 1- \lambda^{(0)} \mu,
\end{equation}
that for the fitting curve ($\mu = \mu^\ast$) yields
numerically\vspace*{-1mm}
\begin{equation}
 1-z_0 \approx \lambda^{(0)} \cdot \mu^\ast
  \sim\, \exp\left(-2.61 \times 10^{122}\right)\!.
                \label{z_0}
\end{equation}\vspace*{-6mm}

\noindent The respective chemical potential~$\tilde{\mu}$ stems
from~(\ref{mu-Chem}):\vspace*{-1mm}
\begin{equation}
\tilde{\mu}_0 =
    -\lambda^{(0)} \mu \cdot k_{\rm B} T\big|_{\mu = \mu^\ast}
    \!\! \sim\,
                - \exp\left(-2.61 \times 10^{122}\right) \cdot k_{\rm B} T.
                                \label{mu-Chem*}
\end{equation}\vspace*{-8mm}

\noindent Besides, from~(\ref{z_0}), (\ref{mu-Chem*}) we
observe\vspace*{-1mm}
\begin{equation}
1-z_0 \approx  - \frac{ \tilde{\mu}_0 }{k_{\rm B} T}.
\end{equation}\vspace*{-4mm}

\noindent \textit{\textbf{Remark.}}\,\,Let us estimate the ratio
$G_2(\mu;z)/G_1(\mu;z)$ for the deformation parameter
value~$\mu=\mu^\ast$ in some $\varepsilon\mu$-neighborhood of the
maximum.\,\,That is, we have $(\lambda^{(0)}-\varepsilon) \mu^\ast <
1-z < (\lambda^{(0)}+\varepsilon)\mu^\ast$.\,\,In addition,
assume~$\varepsilon<1/8$.\,\,Then, making typical estimates we
obtain:\vspace*{-1mm}
\[
G_1(\mu^\ast;z) = \]\vspace*{-8mm}
\[ =\frac52 \bigg\{\! \zeta\bigg(\!\frac52\!\bigg)
 - \zeta\bigg(\!\frac32\!\bigg) (1-\lambda^{(0)}+\varepsilon)\, O(\mu^\ast)
        +  2\pi O(\mu^{\ast3/2})\! \bigg\}\!,       \]\vspace*{-7mm}
        \[
G_2(\mu^\ast;z) =
    G_2(\mu^\ast;z_0) \pm 18 O(\varepsilon) + 12e^3 O(\mu^\ast),
\]

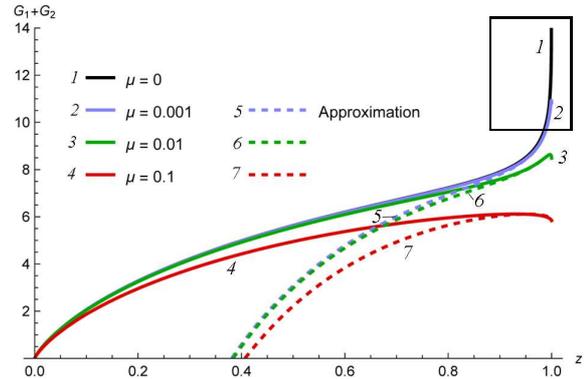
\begin{figure}
\includegraphics[width=75 mm]{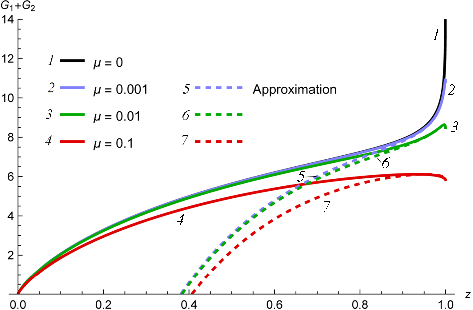}
\begin{picture}(0,0)
\put(-38,93){\framebox(30,42){}}
\end{picture}
\vskip-3mm \caption{``Full-range'' behavior
of~$G_1(\mu;z)+G_2(\mu;z)$ given by solid
curves~\textit{1}--\textit{4}, along with the small~$\mu$
approximation based on~(\ref{G1-Appr}) and~(\ref{G2-Appr}),
presented by dashed curves~\textit{5}--\textit{7}. For
curve~\textit{1}\,-- \ $\mu=0$, for curves~\textit{2} and
\textit{5}\,-- \ $\mu=0.001$, for curves~\textit{3} and
\textit{6}\,-- \ $\mu=0.01$, and for curves~\textit{4},
\textit{7}\,-- \ $\mu=0.1$}\label{fig2}\vspace*{-2mm}
\end{figure}
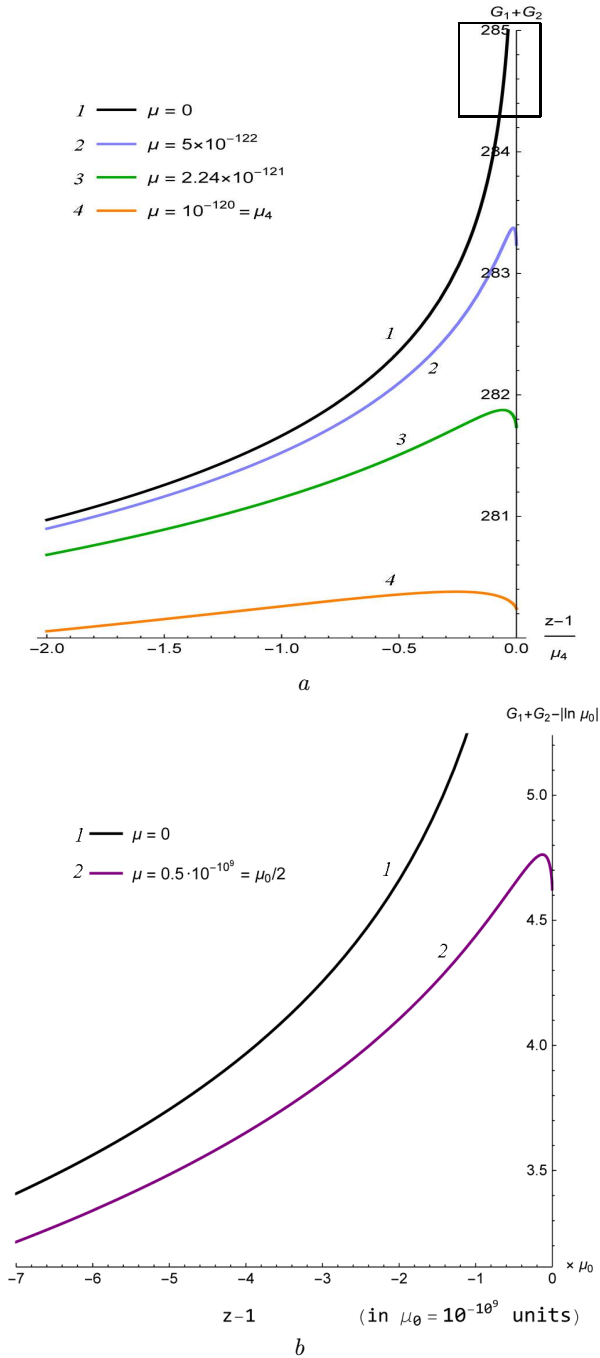
\begin{figure}
\includegraphics[width=72mm,height=87mm]{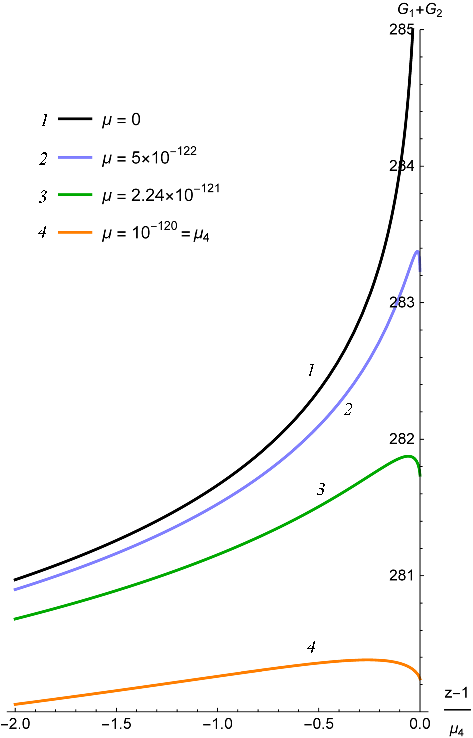}\\
{\it a}\\[1.5mm]
\includegraphics[width=78 mm]{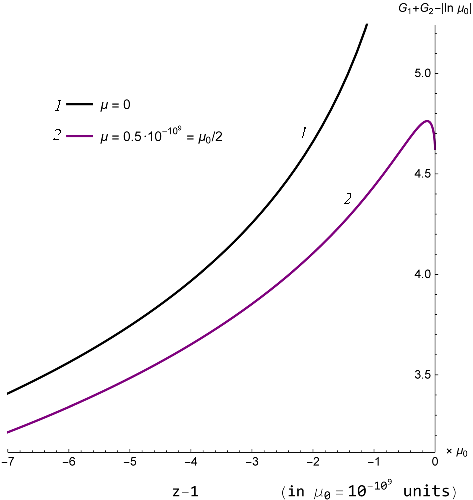}\\
{\it b}
\begin{picture}(0,0)
\put(55,467.5){\framebox(30,35){}}
\end{picture}
\vskip-3mm \caption{\footnotesize
 Dependence of~$G_1(\mu;z)+G_2(\mu;z)$ on~$\delta z \equiv z-1$  measured in
the units of largest  value $\mu_4$, close to~$z=1$ for \mbox{$\mu \sim
10^{-120}$} (\textit{a}).
 Dependence of~$G_1(\mu;z)+G_2(\mu;z)$ on~$\delta z$, shifted
by~$|\ln\left( \mu{_0} \right)|  \approx 2.30$\,$ \times 10^9$, for the
\textit{extremely} small $\mu$-scale given by~$\mu_0 = 10^{-10^9}$,
however, with focus on the extremum position (\textit{b}). This
figure roughly corresponds to the box in the up panel  }\label{fig3}
\end{figure}

\noindent where~$\zeta\bigl(\!\frac32\!\bigr) = 2.61238\,...$,
$\zeta\bigl(\!\frac52\!\bigr) = 1.34149\,...$\,.\,\,Eva\-luating the
ratio within given neighborhood,\vspace*{-1mm}
\[
\frac{G_2(\mu^\ast;z)}{G_1(\mu^\ast;z)} = \frac25
\zeta^{-1}\bigg(\!\frac52\!\bigg) \bigg\{\!
  \bar{G}_2 \pm 18 O(\varepsilon) + \zeta\bigg(\!\frac32\!\bigg)\, \times
 \]\vspace*{-7mm}
 \[
  \times\, 2.05\times\overbrace{ \bar{G}_2 e^{-\bar{G}_2} O(1)}^{
                                O(\mu^{\ast1/2})}
\bigg\}
  \approx  \frac25 \zeta^{-1}\bigg(\!\frac52\!\bigg) \bar{G}_2 \, \pm
 \]\vspace*{-7mm}
\begin{equation}
 \pm \, 8 O(\varepsilon) = 7.79\times 10^{121} \pm 1 \gg 1,
                \label{G2/G1}
\end{equation}
we confirm\,\footnote{
    {Strictly speaking we have to compare also the increments:
    the one for~$G_2(\mu;z)$ in $\mu$-neighbourhood of $z=1$ is of
    the order of~$1$, whilst $G_1(\mu;z)\approx \mathrm{const}$ up to
    $O(\mu)$. So, this should not cause noticeable $z_0$'s shift.}
}
the mentioned possibility to neglect $G_1(\mu;z)$ at
$\Xi_T  \sim 1$.

Now, let us  demonstrate our treatment of~$\mathcal{G}_V(\mu;z)$ by a few
illustrative plots, especially for extremely small~$\mu$.\,\,The
typical picture of~$G_1(\mu;z)+G_2(\mu;z)$ dependence on~$z$ for a
few values of~$\mu$ is shown in Fig.~2.\,\,In addition, the behavior
of small-$\mu$ approximation beyond its validity region (i.e.,~$|\ln
z|\ll 1$), is given for comparison.\,\,The detailed dependencies
from the box in~Fig.~2, i.e., for much more decreased~$\mu$, with
$z$ approaching unity, are visualized by the magnified graphs in
Fig.~\ref{fig3},~\textit{a}, and Fig.~\ref{fig3},~\textit{b}
(extremely large magnification).

{To summarize, the truly deformed case of $\mu > 0$, unlike the
$\mu=0$ (pure Bose gas) case, can produce realistic value of CC, due
to very existence of (extra-tiny) minimum for CC at definite $z_0$
very close to $1$, or due to (very close to the minimum) finite
nonzero value, when we set $z=1$ exactly.}

\section{Discussion and Conclusions}

In this paper, we have derived the $\mu$-deformed extension of
Einstein (gravitational) field equations and studied, in detail, the
dependence  of its term which involves the effective
(i.e.,~$\mu$-dependent) cosmological constant as a function of the
deformation parameter $\mu$  and fugacity $z$.\,\,In effect, we have
shown that, by varying the deformation parameter $\mu$ toward its
extremely small values, it is possible to find the appropriate value
$\mu^\ast$ of $\mu$ that allows us to achieve the ``realistic
value'' of the maximum of the function $\mathcal{G}_V(\mu;z) \equiv
G_{1}(z, \mu) + G_{2}(z, \mu)$ which (through its inverse) provides
the necessary observable value $\Lambda/\mathcal{G}_V(\mu^\ast;z_0)
    \sim 3.8 \times 10^{-123}_{} l_\mathrm{P}^{-2}$ of the effective cosmological
constant $\Lambda/\mathcal{G}_V(\mu;z)$.\,\,Cer\-tain important
points are to be emphasized.

First of all, we stress the very fact of existence of the extremum
of the ECC within the $\mu$-deformed Bose-gas model based approach
to realizing induced gravity.\,\,Clear\-ly this is in contrast to
the usage of conventional (undeformed) Bose gas model.\,\,In\-deed,
within the latter there is no extremum in the corresponding
combination $\Lambda/\mathcal{G}_V(0;z_0)$ of usual $\mu$ = 0
polylogarithms, when $z\rightarrow1$ {and thus no way to get the
value of CC different from zero}.

Second, when analyzing the behavior of ECC, we adopted, without
special justification, that the ratio is
$V/\lambda_{T}^{3}=1$.\,\,This value of the ratio is well-known
marker that indicates the existence of truely quantum effects
\cite{61} and also relates to Bose-like condensate.\,\,Let us
consider this issue in more details, coming unexpectedly to the two
distinct possible situations (and interpretation) of the system
which underlies the induced (entropic) gravity.\,\,These two cases
of inducing system correspond to principally different scales,
namely (A) the scale (size) of dwarf galaxies and (B) the scale of
observable Universe.

{(A)} In this case, the ratio $V/\lambda_{T}^{3}$ is of the order of
unity due to the fact that both the scale of dwarf galaxies and the
thermal wavelength scale of the dark matter particles are few kpc
(see, e.g., \cite{18}) if the $\mu$-deformed fuzzy dark matter (DM)
model is adopted, with the mass of (Bose-like condensate) ultralight
DM particles being $m = 10^{-22}$~eV.

{(B)} In this case, the ratio $V/\lambda_{T}^{3}$ is of the order of
unity due to the fact that both the scale of observable Universe and
the thermal wavelength  scale  are of the order $\sim$$2.8 \times
10^{7}$~kpc if the particles, forming \cite{63} Bose-like condensate
DM, are adopted to be ``massive gravitons'' (it is reasonable to
term this \textit{pre-gravitons} in view of their role in generating
induced gravity) with mass $m = 10^{-36} $--$ 10^{-33}$~eV, see
\cite{62}.\,\,Note that, in the context of DM of dwarf galaxies, the
Bose condensate of massive gravitons was considered in
\cite{63}.\,\,In our treatment, we deal with condensate of massive
pre-gravitons, i.e.\,\,viewed as  the system which induces real
gravity along with effective cosmological constant.\,\,In connection
with this picture of inducing CC, it is worth to mention the work of
R.~Garattini and M.~Faizal in which it was shown that the appearance
of  cosmological constant is strictly related with just the
deformation of Generalized Uncertainty Principle being behind the
deformed Wheeler--DeWitt equation, see~\cite{64}.

The real physical nature of the parameter $\mu$ of
\mbox{$\mu$-deformed} inducing system, like that of dark matter, is
unknown.\,\,Ho\-we\-ver the significance or meaning of its use can
be justified: the deformation enables to take into account
(possible) compositeness of particles and/or their interactions~--
natural reasons to deal with quantum system strongly deviating from
ideal Bose gas.\,\,This reason is quite similar to the usage of the
deformed Bose gas model in another context~-- for (successful)
modelling the properties of quantum correlations of pions occurring
in heavy-ion collisions at RHIC, see Ref.~\cite{15} and also the
paragraph after Eq.~(\ref{S(mu)}).\,\,Ho\-we\-ver, explicit physical
content of $\mu$ in the both two cases, of ultralight dark matter
(as Bose-like condensate) i.e. the galactic one, and the observed
Universe case, may be different since the particles (and their
constituents) do differ -- ultralight ``bosons'' versus
``pre-gravitons''.

Let us stress that our approach to ``explain'' the issue of
observable value of CC grounds on two pillars: deformation and
(macroscopic) quantumness of underlying or inducing system.\,\,It is
also worth to note the following: the fact that the issue of CC in
our treatment turns out to be related with the concept of DM is
quite alike the joint treatment of DM and dark energy, appearing in
another contexts in some papers, see e.g.~\cite{65,66}.

As our final remark let us mention that the analysis of $\mu$-deformed Friedmann
equations stemming from the $\mu$-deformed Einstein equations derived and
studied herein, will be published in a separate  paper.

\vskip2mm \textit{The authors acknowledge anonymous reviewers for
valuable comments and remarks which helped to improve the
manuscript. This research was funded by the National Academy of
Sciences of Ukraine by its priority project No.\,0122U000888 and was
supported by the Simons Foundation grant~1290587.}


\end{document}